# On the maximum efficiency of realistic heat engines

**E. N. Miranda**
Facultad de Ingeniería
Universidad de Mendoza
5500, Mendoza – Argentina

Área de Ciencias Exactas
CRICYT – CONICET
5500, Mendoza - Argentina

Departamento de Física
Universidad Nacional de San Luis
5700, San Luis – Argentina

**Abstract:** In 1975, Courzon and Ahlborn studied a Carnot engine with thermal losses and got an expression for its efficiency that described better the performance of actual heat machines than the traditional result due to Carnot. In their original derivation, time appears explicitly and this is disappointing in the framework of classical thermodynamics. In this note a derivation is given without any explicit reference to time.

The efficiency of thermal machines has attracted the attention of researchers since the nineteenth century. The problem is interesting both from a conceptual point of view and for obvious practical reasons. It is well known that for an ideal heat engine working between a hot reservoir at $T_h$ and a cold one at $T_c$, the maximum efficiency $\eta$ is given by the famous result due to Carnot: $\eta = 1 - (T_c/T_h)$; however, the efficiency of real heat engine is considerable less than that.

In 1975, Courzon and Ahlborn [1] studied an ideal Carnot engine in contact with two reservoirs but the heat conduction process was not ideal; there were heat resistances different from zero between the reservoirs and the engine. Those authors introduced time explicitly in their analysis and got an interesting result: the efficiency of the system came out to be $\eta_{CA} = 1 - (T_h/T_c)^{1/2}$ and had no temporal dependence. Moreover, a comparison between the Courson-Ahlborn efficiency and those of real heat engines showed good agreement. Later, Rebhan [2] studied a Carnot engine with thermal losses and friction. He found that the efficiency of these machines was bounded from above by the Courzon-Ahlborn expression. Recently, van den Broeck [3] has shown, in the framework of non-equilibrium thermodynamics, that the efficiency of any thermal engine is bounded from above by $\eta_{CA}$. This suggests that the result of Courson and Ahlborn has a deeper meaning than that previously thought. However, there is one point that is disappointing at least for the present author: time is explicitly used in the original deduction of $\eta_{CA}$. For this reason, in this note a deduction of the Courzon-Ahlborn efficiency is given without any reference to time.

Assume a Carnot heat engine in contact with two heat reservoirs at $T_h$ ($T_c$) respectively as shown in Figure 1. There is a thermal resistance $R_h$ ($R_c$) between the hot (cold) reservoir and the engine. Let call $Q_h$ ($Q_c$) the heat that flow from (to) the heat (cold) reservoir and $W$ the work output. The heat engine is actually working between a high temperature $T_1$ and a cold one $T_2$, given by:

$$T_1 = T_h - Q_h R_h$$

$$\begin{aligned}T_2 &= T_c + Q_c R_c \\ &= T_c + (Q_h - W) R_c\end{aligned} \quad (1)$$

The efficiency of this Carnot engine is:

$$\eta = 1 - \frac{T_2}{T_1} = \frac{W}{Q_h} \quad (2)$$

If the values given by equation (1) are used, the work $W$ comes out to be:

$$W = \frac{T_h - T_c - Q_h(R_h + R_c)}{T_h - Q_h(R_h + R_c)} Q_h \quad (3)$$

The useful work from the thermal machine comes out to depends on the heat flux from the hot reservoir and the total heat resistance $R = R_h + R_c$. To obtain the maximum available work from the machine, one should maximise (3) respect to $Q_h$. It is found that the work is a maximum when the heat is given by:

$$Q_{max} = \frac{T_h - \sqrt{T_h T_c}}{R} \tag{4}$$

And the maximum work results:

$$W_{max} = \frac{1}{R\sqrt{T_c T_h}} \left(T_h - \sqrt{T_c T_h}\right)\left(\sqrt{T_c T_h} - T_c\right) \tag{5}$$

From (4) and (5) one concludes that the maximum efficiency of the heat engine is:

$$\eta_{max} = 1 - \sqrt{\frac{T_c}{T_h}} \tag{6}$$
$$= \eta_{CA}$$

This calculation shows that the efficiency proposed by Courson and Ahlborn is the maximum one attainable by a Carnot engine when there are heat resistances between the machine and the reservoirs. In our deduction nor the time that the working fluid is in contact with the reservoirs neither the cycle velocity appear. In this sense, the present way of getting $\eta_{CA}$ is more adequate for a conventional course in classical thermodynamics.

**Acknowledgement:** The author is staff researcher of the National Research Council of Argentina (CONICET).

# Figure caption

**Figure 1:** This figure shows the system analyzed in the text. An ideal Carnot engine is in contact with two reservoirs at temperature $T_h$ and $T_c$ respectively. However, there are thermal resistances between the machine and the reservoirs that are indicated as $R_h$ and $R_c$. Consequently, the engine works between temperatures $T_1$ and $T_2$. $Q_h$ is the heat that flows from the hot reservoir to the engine, $W$ is the work given by the engine and $Q_c$ is the heat that goes toward the cold reservoir. The aim of this note is to evaluate the maximum efficiency of the system.

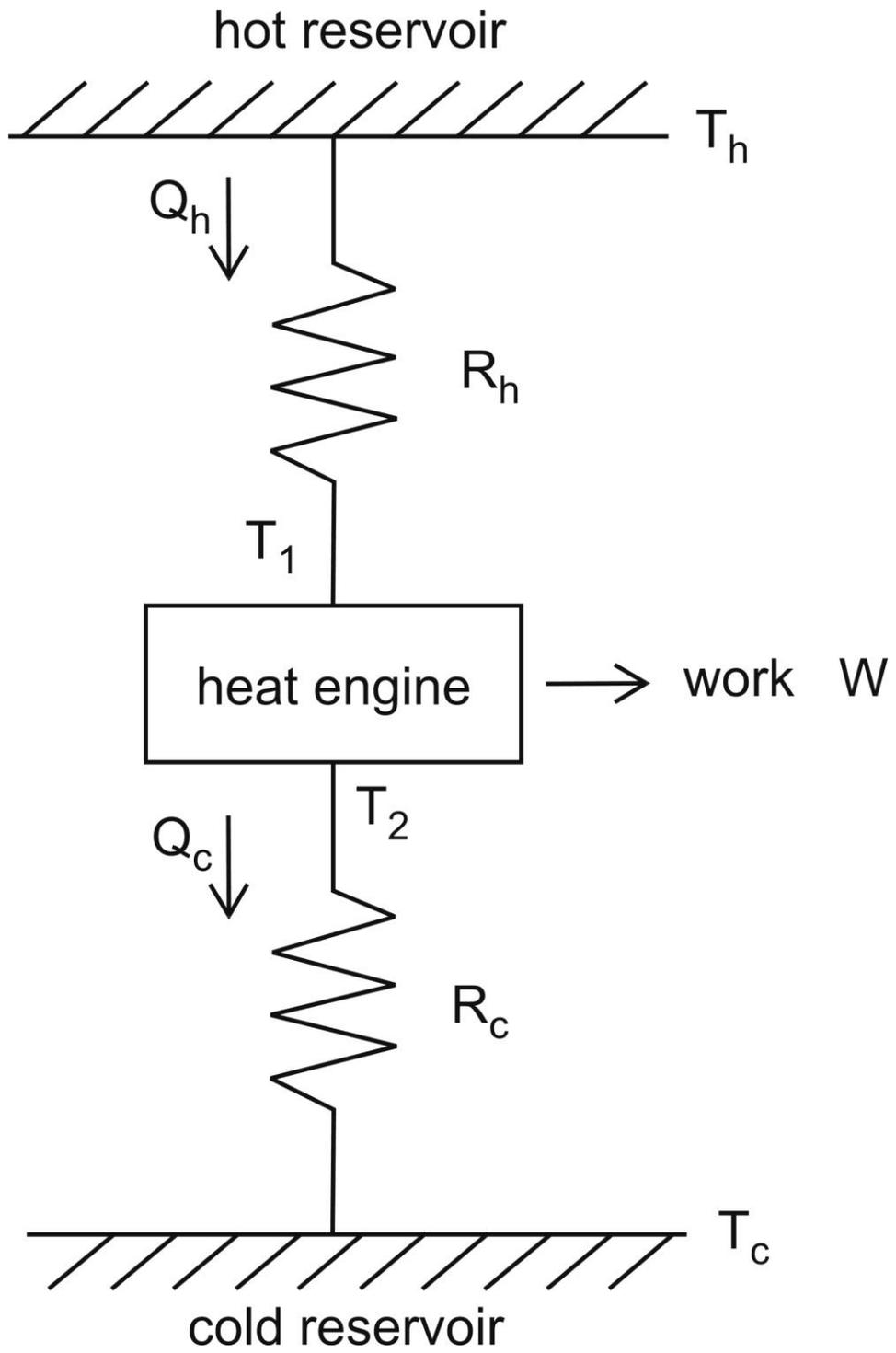

**Figure 1**